\documentclass[twocolumn,10pt]{article}

\usepackage[a4paper, total={6.8in, 9.5in}]{geometry}

\usepackage{graphicx} 
\graphicspath{{figures/}} 
\usepackage{float}
\usepackage{amsmath}
\usepackage{esint} 
\usepackage{mathrsfs} 

\usepackage{bm} 

\usepackage{indentfirst} 

\usepackage{xcolor}
\usepackage{doi}

\usepackage[square,numbers]{natbib}

\usepackage{hyperref}
\usepackage{authblk}
\hypersetup{
	bookmarksnumbered=true,
    colorlinks=true, 
    linktoc=page,     
    linkcolor=blue,  
    citecolor=teal,
    urlcolor=teal
}

\title{\textbf{Analytic Hall Magnetohydrodynamics toroidal equilibria via the energy-Casimir
variational principle}}
\author[1]{A. Giannis\thanks{}}
\author[1,2]{D. A. Kaltsas\thanks{}}
\author[1]{G. N. Throumoulopoulos\thanks{}}
\affil[1]{Department of Physics, University of Ioannina, Ioannina, Greece, GR 451 10 }
\affil[2]{Department of Physics, International Hellenic University, Kavala, Greece, GR 654 04 }
\date{}

\begin{document}

\twocolumn[
  \begin{@twocolumnfalse}

    \maketitle
    \begin{abstract}
    Equilibrium equations for magnetically confined, axisymmetric plasmas are derived by means of the energy-Casimir variational principle in the context of Hall magnetohydrodynamics (MHD). This approach stems from the noncanonical Hamiltonian structure of Hall MHD, the simplest, quasineutral two-fluid model that incorporates contributions due to ion Hall drifts. The axisymmetric Casimir invariants are used, along with the Hamiltonian functional to apply the energy-Casimir variational principle for axisymmetric two-fluid plasmas with incompressible ion flows. This results in a system of equations of the Grad-Shafranov-Bernoulli (GSB) type with four free functions. Two families of analytic solutions to the GSB system are then calculated, based on specific choices for the free functions. These solutions are subsequently applied to Tokamak-relevant configurations using proper boundary shaping methods. The Hall MHD model predicts a departure of the ion velocity surfaces from the magnetic surfaces which are frozen in the electron fluid. This separation of the characteristic surfaces is corroborated by the analytic solutions calculated in this study. The equilibria constructed by these solutions exhibit favorable characteristics for plasma confinement, for example they possess closed and nested magnetic and flow surfaces with pressure profiles peaked at the plasma core. The relevance of these solutions to laboratory and astrophysical plasmas is finally discussed, with particular focus on systems that involve length scales on the order of the ion skin depth.
    \end{abstract}
\vspace{1cm}
  \end{@twocolumnfalse}
] 

{
  \renewcommand{\thefootnote}%
    {\fnsymbol{footnote}}
  \footnotetext[1]{a.giannis@uoi.gr}
  \footnotetext[2]{d.kaltsas@uoi.gr}
 \footnotetext[3]{ gthroum@uoi.gr}
 
}

\section{Introduction}\label{sec:1}
Hall Magnetohydrodynamics (Hall MHD) is the simplest MHD model that incorporates two-fluid effects and is obtained by consistently reducing the complete two-fluid equations of motion, under the assumptions of quasineutrality and vanishing electron inertia. The non-dimensional Hall MHD equations read (in Alfvén units) as: 
\begin{align}
  &\bm{E}+\bm{v}\times\bm{B}=d_i(\bm{J}\times\bm{B}-\bm{\nabla} \cdot\textbf{P}_e), \label{eq:ohm} \\
  &\partial_t \rho = -\bm{\nabla}\cdot(\rho\bm{v}), \label{eq:continuity} \\
  &\partial_t\bm{v}=-\bm{\nabla}\cdot\left(\textbf{P}+\frac{|\bm{v}|^2}{2} \textbf{I}\right)+\bm{v}\times(\bm{\nabla}\times\bm{v})+\bm{J}\times\bm{B}, \label{eq:momentum} \\
  &\partial_t\bm{B}=\bm{\nabla}\times \left[\bm{v}\times\bm{B}-d_i(\bm{J}\times\bm{B})\right], \label{eq:induction}
\end{align}
where $\bm{E}$, $\bm{B}$ are the electric and magnetic fields respectively, $\bm{v}$ is the ion fluid velocity, $\rho$ is the total mass density, $\bm{J}=\bm{\nabla}\times\bm{B}$ is the current density, $\textbf{P}$ is the total plasma pressure tensor and $\textbf{P}_e$ is the electron pressure tensor, $\textbf{I}$ is the identity tensor, $d_i=c/({\omega_p}_i L)$ is the ion skin depth, normalized by a characteristic length scale $L$, $c$ is the speed of light, and ${\omega_p}_i$ is the ion plasma frequency. Equation \eqref{eq:ohm} is a generalized Ohm's law resulting from  the electron momentum equation for vanishing electron mass and infinite electrical conductivity, Eq.~\eqref{eq:continuity} is the continuity equation, Eq.~\eqref{eq:momentum} is the momentum equation, while Eq.~\eqref{eq:induction} is an induction equation for the magnetic field, resulting from combining Ohm's law \eqref{eq:ohm} with Faraday's law $\bm{\nabla}\times\bm{E}=-\partial_t \bm{B}$. The right-hand side of Eq.~\eqref{eq:ohm} contains the so-called Hall term, as well as the electron pressure term. These terms cause a detachment of the ion fluid from the electron one, in length scales comparable or smaller than the ion skin depth.

For the study of most laboratory and astrophysical plasma systems, the ideal MHD model is usually being employed, ignoring terms that scale as $d_i$ in Eqs.~\eqref{eq:ohm}--\eqref{eq:induction} and thus eliminating two-fluid effects. This one-fluid description is an adequate framework for the study of large time and length scale processes i.e. dynamical phenomena with time-scales larger than the ion cyclotron frequency and plasma structures with length scales much larger than the ion skin depth. However, some fundamental processes cannot be described within this framework as they originate from the two-fluid nature of the plasmas, e.g. fast magnetic reconnection, various micro-instabilities that trigger or enhance turbulence and transport and wave modes that are not present in a single fluid framework. On top of that, there exist laboratory plasma systems that involve processes and structures with length scales comparable to the ion skin depth, for example magnetically confined plasmas with current sheets, thin boundary layers, and steep gradients, such as Tokamak plasmas  with a steep pressure gradient in the edge region and pressure pedestals that develop in the transition to improved confinement regimes (L-H transition) \cite{wagner_1982, wagner_2007}, and neoclassical diffusion \cite{stroth}. The development of a theory for the pressure pedestals in high (H) confinement mode in Tokamaks in \cite{Guzdar_2005}, which was based on double-Beltrami Hall MHD equilibrium states, is indicative of the relevance of Hall MHD with H-mode plasmas (see also \cite{Yoshida_2001}). It has also been established that Hall effects are relevant to the so-called tearing mode instability \cite{zhang_2017} that occurs in both laboratory and astrophysical plasmas. To further acknowledge the importance of the Hall MHD model in studying the magnetic confinement of plasmas, we stress that the omission of the Hall term in Tokamaks has been previously criticised in \cite{gourdain}. As concerns astrophysical plasmas, there are many systems that can be described in the framework of Hall MHD, with the most notable example possibly being the corroboration by in situ satellite measurements that magnetic reconnection in Earth's magnetosphere is described by a two-fluid model \cite{berkowitz}. The usefulness of Hall MHD is not limited to the above examples; the interested reader is referred to \cite{mininni} for even more examples of astrophysical systems described by Hall MHD, like dense molecular clouds, protostellar disks and neutron stars, to name only a few.

The present paper addresses the construction of analytic, axisymmetric, Hall MHD equilibrium states for Tokamak plasmas with shaped boundaries. The starting point of this construction is the derivation of the Hall MHD equilibrium equations by exploiting the energy-Casimir principle, a Hamiltonian variational principle that stems from the noncanonical Hamiltonian structure of Hall MHD. Then the equilibrium equations are cast in a the form of a Grad-Shafranov system and we provide two special analytic solutions which are used to construct Tokamak-pertinent equilibria. This work is organized as follows: In Section \ref{sec:2} we deduce equilibrium equations for axisymmetric two-fluid plasmas with incompressible ion flows in light of the Energy-Casimir variational principle. The resulting Grad-Shafranov-Bernoulli system is then solved analytically for two specific cases in Section \ref{sec:3}. In Section \ref{sec:4} we apply these solutions to up-down symmetric, International Thermonuclear Experimental Reactor (ITER)-like configurations, while in Section \ref{sec:5} we summarize and discuss the results of the present study.

\section{Energy-Casimir equilibrium variational principle}\label{sec:2}

It has been established (e.g. \cite{holm_1987,lingam_2015}) that Eqs.~\eqref{eq:continuity}-\eqref{eq:induction} possess a noncanonical Hamiltonian structure \cite{morrison}, in the sense that the system's dynamics can be described by a set of generalized Hamilton's equations
\begin{equation}\label{eq:generalized_Hamilton}
  \partial_t \bm{\eta} = \{\bm{\eta},\mathcal{H}\}\,,
\end{equation}
where $\bm{\eta}=(\rho,\bm{v},\bm{B})$ represents the dynamical variables of the model, $\mathcal{H}=\mathcal{H}[\bm{\eta}]$ is the Hamiltonian functional, an integral over the plasma volume, and $\{F,G\}$ is a noncanonical Poisson bracket, which is bilinear and antisymmetric and also satisfies the Jacobi identity \cite{morrison}. The noncanonical Poisson bracket of Hall MHD has been found in \cite{holm_1987,lingam_2015}. Such brackets can be degenerate, i.e. there exist functionals $C$ that Poisson-commute with any other functional $F$ of the dynamical variables, that is
\begin{equation}\label{eq:casimir_commutation}
  \{C,F\}=0\,, \quad \forall F.
\end{equation}
The functionals $C$ are thus invariants of the system, called Casimirs or Casimir invariants \cite{morrison}. Regarding equilibrium points $\bm{\eta}_e$ of the Hamiltonian system, these can be calculated by 
\begin{eqnarray}
    \{\bm{\eta}_e,\mathcal{H}\}=0\,. \label{equilibrium_points}
\end{eqnarray}
However, as $C$ satisfy \eqref{eq:casimir_commutation}, the functional $\mathcal{H}$ in \eqref{equilibrium_points} can be replaced by a generalized Hamiltonian formed by a linear combination of the Hamiltonian $\mathcal{H}$ and the Casimirs $C$. Then, nonsingular equilibrium points satisfy
\begin{equation}\label{eq:energy_casimir}
  \delta\left(\mathcal{H}-\sum_iC_i\right)[\bm{\eta}_e]=0\,,
\end{equation}
Relation \eqref{eq:energy_casimir} is an expression of the so-called Energy-Casimir variational principle, and is essentially a sufficient condition for equilibrium \cite{morrison}.

Here, we apply this principle considering a two-fluid plasma with massless electrons, homogeneous ion density and anisotropic electron pressure. The equations of motion are given by \eqref{eq:ohm}--\eqref{eq:induction} with $\textbf{P} = P_i +\textbf{P}_e$. Two additional equations are needed to close the system, one for the ion pressure $P_i$ and one for the electron pressure $\textbf{P}_e$.  We have shown in \cite{Kaltsas_2021} that the electron pressure equation in the Hall MHD limit becomes \cite{Hunana_2019a}
\begin{eqnarray}
    \textbf{B} \times \textbf{P}_e +(\textbf{B}\times \textbf{P}_e)^T =0\,, \label{electron_pressure_eq}
\end{eqnarray}
which is solved by gyrotropic pressure tensors of the form
\begin{eqnarray}
    \textbf{P}_e = \frac{P_{e\parallel} - P_{e\perp}}{B^2} \textbf{B}\textbf{B} +P_{e\perp} \textbf{I}= \sigma \textbf{B}\textbf{B}+P_{e\perp} \textbf{I}\,, \label{gyrotropic_press_tensor}
\end{eqnarray}
where $\bm{B}\bm{B} = B_iB_j$ is the tensor product of $\bm{B}$ with itself, $P_{e\parallel}$ and $P_{e\perp}$ denote the electron pressures parallel and perpendicular to the magnetic field, respectively and we defined a function $\sigma := (P_{e\parallel}-P_{e\perp})/B^2$ which quantifies the anisotropy of the electron pressure. Evidently, for $\sigma=0$ the electron pressure is isotropic.

As regards the closure equation for the ion pressure, this is replaced by plasma incompressibility to simplify the subsequent analysis. This means that the mass density is assumed to be homogeneous $(\rho=1)$ throughout the plasma volume and thus the fluid velocity is divergence-free $\bm{\nabla}\cdot\bm{v}=0$. This can be imposed as a constraint in the energy-Casimir variational principle \eqref{eq:energy_casimir} as follows
\begin{eqnarray}
    \delta \left(\mathcal{H} - \sum_i C_i + \int d^2x \, P_i ln(\rho)\right)=0 \,. \label{eq: energy_casimir_constr}
\end{eqnarray}
where the ion pressure $P_i$ is introduced as a Lagrange multiplier. An analogous technique for introducing the pressure in the Lagrangian description of incompressible ideal MHD has been proposed by Newcomb in \cite{Newcomb_1962}.

The Hamiltonian for axisymmetric Hall MHD plasmas with electron pressure anisotropy reads in the standard cylindrical coordinates $(R,\phi,Z)$, as
\begin{eqnarray}\label{eq:hamiltonian}
  \mathcal{H}=\int\limits_\mathcal{S} d^2x \ \bigg(\rho \frac{v_\phi^2}{2} + \frac{B_\phi^2}{2}  + \frac{|\bm{\nabla}\Psi|^2}{2R^2}  \nonumber \\
+ \rho\frac{|\bm{\nabla}X|^2}{2R^2}   + \rho U_e(\rho,|\textbf{B}|) \bigg),
\end{eqnarray}
where $\Psi$ and $X$ are the two poloidal flux functions for the magnetic and the velocity fields, respectively, and $\mathcal{S}$ is the plasma cross section normal to the $\phi$ direction with closed boundary $\partial \mathcal{S}$. Note that the ion internal energy is not included in $\mathcal{H}$ as it has been taken into account through the constraint introduced in \eqref{eq: energy_casimir_constr}. The axisymmetric magnetic and velocity fields are given in terms of $\Psi$ and $X$ by the following relations
\begin{align}
     \bm{B} &= RB_\phi \bm{\nabla} \phi + \bm{\nabla} \Psi \times \bm{\nabla} \phi ,\label {eq:poloidal_rep_B}\\
    \bm{v} &= Rv_\phi \bm{\nabla} \phi  + \bm{\nabla} X \times \bm{\nabla} \phi. \label{eq:poloidal_rep_v}    
\end{align}

In Eq.~\eqref{eq:hamiltonian}, $U_e$ is the electron internal energy function, which should satisfy \cite{Morrison_1982}:
\begin{eqnarray}
    \frac{\partial U_e}{\partial \rho} = \frac{P_{e\parallel}}{\rho^2} \,, \label{U_e_rho}\\
    \frac{\partial U_e}{\partial |\textbf{B}|} = - \frac{\sigma}{\rho}|\textbf{B}|\,. \label{U_e_|B|}
\end{eqnarray}

The Hall MHD Casimir invariants with axial and helical symmetry have been previously calculated in \cite{kaltsas_2018} and the recovery of the corresponding invariants in the MHD limit has been discussed in \cite{Kaltsas_2017}. The axisymmetric Hall MHD Casimirs are
\begin{align}
  &C_1 = \int\limits_\mathcal{S} d^2x \ (R^{-1}B_\phi + d_i\Omega)\mathcal{F}(\Phi), \label{eq:casimir_1} \\
  &C_2 = \int\limits_\mathcal{S} d^2x \ R^{-1}B_\phi\mathcal{G}(\Psi), \label{eq:casimir_2} \\
  &C_3 = \int\limits_\mathcal{S} d^2x \ \rho\mathcal{M}(\Phi), \label{eq:casimir_3} \\
  &C_4 = \int\limits_\mathcal{S} d^2x \ \rho\mathcal{N}(\Psi), \label{eq:casimir_4}
\end{align}
where $\mathcal{F}$, $\mathcal{G}$, $\mathcal{M}$ and $\mathcal{N}$ are free functions, i.e. arbitrary functions of their arguments; $\Omega=(\bm{\nabla}\times\bm{v_p})\cdot \bm{\nabla}\phi=-\frac{1}{R^2}\Delta^\star X$, with $\bm{v_p}=\bm{\nabla}X\times\bm{\nabla}\phi$ being the poloidal component of the velocity field and $\Delta^\star\equiv R^2\bm{\nabla}\cdot(\bm{\nabla}/R^2)$ is the so-called Shafranov operator. In Eqs.~\eqref{eq:casimir_1}--\eqref{eq:casimir_4}, $\Phi$ is a generalized flux function defined as $\Phi=\Psi+d_iR v_\phi$. The physical interpretation of these symmetric Casimirs has been discussed in previous studies (e.g. \cite{Kaltsas_2017}), however, for completeness we mention that $C_1$ and $C_2$ are related to the generalized Hall MHD cross-helicity and the magnetic helicity, respectively, while $C_3$ and $C_4$ are associated with the conservation of mass, toroidal angular momentum and magnetic flux. 

Employing the constrained energy-Casimir variational principle \eqref{eq: energy_casimir_constr} leads to the following set of Euler-Lagrange equations
\begin{eqnarray}
    &&\hspace{-5mm}\delta  P_i : \; \rho =1\,, \label{delta_Pi}\\ 
    &&\hspace{-5mm}\delta \rho : \; \frac{P_i + P_{e\parallel}}{\rho}  + U_{e} = \mathcal{M}(\Phi) +\mathcal{N}(\Psi) - \frac{v^2}{2}\,, \label{delta_rho}\\
    &&\hspace{-5mm}\delta X: \;   d_i \bm{\nabla}\cdot\left(\frac{\bm{\nabla}\mathcal{F}}{R^2}\right) = \bm{\nabla}\cdot\left(\rho \frac{\bm{\nabla} X}{R^2}\right)\,, \label{delta_X}\\
    && \hspace{-5mm} \delta v_\phi : \; \rho v_\phi = d_i R (R^{-1}B_\phi +d_i\Omega)\mathcal{F}' \nonumber \\
    && \hspace{15mm}+d_i R \rho \mathcal{M}'(\Phi)\,. \\
    && \hspace{-5mm} \delta B_\phi: \; B_\phi = \frac{\mathcal{F}(\Phi) + \mathcal{G}(\Psi)}{R^2(1-\sigma)}\,, \\
    && \hspace{-5mm} \delta \psi : \; \bm{\nabla} \cdot\left[(1-\sigma) \frac{\bm{\nabla} \Psi}{R^2}\right]+ R^{-1} B_\phi (\mathcal{F}' + \mathcal{G}')\nonumber \\
    && \hspace{5mm} +d_i \Omega \mathcal{F}'(\Phi) +\rho \mathcal{N}'(\Psi) +\rho \mathcal{M}'(\Phi)=0\,.
\end{eqnarray}
After some manipulations the system above can be cast in a Grad-Shafranov-Bernoulli (GSB) form

\begin{equation}\label{eq:first_gs}
    \begin{aligned}
    d_i^2\mathcal{F}'(\Phi)R^2\bm{\nabla}\cdot & \left[\mathcal{F}'(\Phi)\frac{\bm{\nabla}\Phi}{R^2}\right]-\frac{[\mathcal{F}(\Phi)+\mathcal{G}(\Psi)]\mathcal{F}'(\Phi)}{1-\sigma} \\
    &-\mathcal{M}'(\Phi)R^2+\frac{\Phi-\Psi}{d_i^2}=0,
    \end{aligned}
\end{equation}

\begin{eqnarray}\label{eq:second_gs}
  \bm{\nabla}\cdot\left[(1-\sigma)\frac{\bm{\nabla}\Psi}{R^2}\right]+\frac{[\mathcal{F}(\Phi)+\mathcal{G}(\Psi)]\mathcal{G}'(\Psi)}{1-\sigma}\nonumber\\
  +\mathcal{N}'(\Psi)R^2 + \frac{\Phi-\Psi}{d_i^2}=0,
\end{eqnarray}

\begin{equation}\label{eq:pressure}
  P=P_i+P_e=\mathcal{M}(\Phi)+\mathcal{N}(\Psi) - \frac{v^2}{2}.
\end{equation}
 Relations \eqref{eq:first_gs} and \eqref{eq:second_gs} are equations of the Grad-Shafranov type (\cite{grad_1958,shafranov}) determining the poloidal flow function $\Phi$ (it can be deduced from \eqref{delta_X} that $\Phi$ is associated with the poloidal ion flow) and the poloidal magnetic flux function $\Psi$. Finally, relation \eqref{eq:pressure} is a Bernoulli equation that determines the total plasma pressure and is independent of the other two equations as a result of the incompressibility assumption. Similar results have been obtained by standard methods for reducing the generic equilibrium equations to a GSB system in \cite{Throumoulopoulos_2006}, while as is well known, for compressible plasmas all three equations are coupled to each other (e.g. see \cite{kaltsas_2018, Kaltsas_2017, Hameiri_2013, Guazzotto_2015}).

Henceforward we assume isotropic electron pressure, i.e. $\sigma =0$ in order to be able to find analytic solutions to the GS system \eqref{eq:first_gs}--\eqref{eq:second_gs}.  In this case, from Eqs.~\eqref{U_e_rho}--\eqref{U_e_|B|} we deduce that $U_e$ is a function of $\rho$ only and thus the electron pressure is constant since $\rho=1$. Additionally, a specific choice for the four arbitrary functions $\mathcal{F},\mathcal{G},\mathcal{M},\mathcal{N}$ should be made for the derivation of analytic solutions. More specifically, we will adopt the following general ansatz for the free functions
\begin{align}
  \mathcal{F}(\Phi)&=f_0+f_1\ \Phi, \label{eq:ansatz1} \\
  \mathcal{G}(\Psi)&=g_0+g_1 \ \Psi, \label{eq:ansatz2} \\
  \mathcal{M}(\Phi)&=m_0+m_1\Phi+\frac{1}{2}m_2\Phi^2, \label{eq:ansatz3} \\
  \mathcal{N}(\Psi)&=n_0+n_1\Psi +\frac{1}{2}n_2\Psi^2, \label{eq:ansatz4}
\end{align}
 where $f_0, \ f_1, \ g_0, \ g_1, \ m_0, \ m_1, \ m_2, \ n_0, \ n_1, \ n_2$ are free, constant parameters, which are determined in view of specific equilibrium characteristics. We note that $f_0$ and $g_0$ are related to the vacuum toroidal magnetic field, hence a selection such that $f_0 + g_0\neq 0$ results in a Tokamak-relevant $1/R$-vacuum field, while a selection for which $f_0+g_0 = 0$ could describe Spheromak-relevant equilibria. The rest of the parameters $f_1, \ g_1, \ m_1, \ m_2, \ n_1, \ n_2$ are associated with the self-consistent fields and plasma quantities. Below we examine two notable cases of analytic equilibria; the first corresponds to $m_2=n_2=0$ and the second to $m_2\neq 0$, $n_2\neq 0$.

\section{Analytic solutions to the GS system}\label{sec:3}

\subsection{Double Beltrami equilibria}

The coupled system of GS equations resulting from \eqref{eq:first_gs}--\eqref{eq:second_gs} with \eqref{eq:ansatz1}--\eqref{eq:ansatz4} and $m_2=n_2=0$ can be cast in the following matrix form
\begin{equation}\label{eq:DB_matrix_nonhomogeneous}
	\Delta^\star \begin{pmatrix}
		\Psi \\
		X
	\end{pmatrix}=\begin{pmatrix}
		\mathcal{W}_1 & \mathcal{W}_2 \\
		\mathcal{W}_3 & \mathcal{W}_4
	\end{pmatrix}
	\begin{pmatrix}
		\Psi \\
		X
	\end{pmatrix} + \begin{pmatrix}
		A_1 \\
		A_2
	\end{pmatrix} + \begin{pmatrix}
		B_1 \\
		B_2
	\end{pmatrix}R^2,
\end{equation}
where we defined 
\begin{equation}\label{eq:matrix_coefficients}
    \begin{split}
        &\mathcal{W}_1=- \left(g_1^2-\frac{1}{d_i^2}\right), \ \mathcal{W}_2=- \left(\frac{g_1}{d_i} + \frac{1}{f_1 d_i^3}\right), \\
        &\mathcal{W}_3=\left(\frac{g_1}{d_i} + \frac{1}{f_1d_i^3}\right), \ \mathcal{W}_4=\left(\frac{1}{d_i^2}-\frac{1}{f_1^2d_i^4}\right),
    \end{split}
\end{equation}
and 
\begin{equation}\label{eq:DB_A_i_coefficients}
    \begin{aligned}
        &A_1 = - \left(g_0g_1 - \frac{f_0}{d_i^2 f_1}\right), \quad A_2 = \left(f_1g_0 + \frac{f_0}{d_i^2f_1}\right),\\
        &\quad B_1 = -n_1 , \quad B_2 = \frac{m_1}{d_i f_1}.
    \end{aligned}
\end{equation}
The above matrix coefficients \eqref{eq:matrix_coefficients}, \eqref{eq:DB_A_i_coefficients} depend solely on the ansatz parameters (see eqs. \eqref{eq:ansatz1}--\eqref{eq:ansatz4}) and the ion skin depth $d_i$.

The homogeneous part of \eqref{eq:DB_matrix_nonhomogeneous} (i.e. $A_1=A_2=B_1=B_2=0$) can also be deduced by considering fields  $\bm{\mathcal{V}}=(\bm{B}_h,\bm{v}_h)$ that satisfy
\begin{equation}\label{eq:db}
  \bm{\nabla}\times\bm{\nabla}\times\bm{\mathcal{V}}+b_1(\bm{\nabla}\times\bm{\mathcal{V}})+b_2\bm{\mathcal{V}}=0,
\end{equation}
where
\begin{equation}\label{eq:DB_b1,2_definitions}
    b_1=-g_1-\frac{1}{f_1d_i^2}, \quad \quad b_2=\frac{f_1+g_1}{f_1d_i^2}\,.
\end{equation}

Equation \eqref{eq:db} can be solved by appropriate superpositions of Beltrami fields $\bm{\mathcal{A}}_\pm$, i.e. by eigenvectors of the curl operator \cite{hudson_2007, bunity_2014}
\begin{equation}\label{eq:beltrami}
  \bm{\nabla}\times\bm{\mathcal{A}}_\pm=\lambda_\pm\bm{\mathcal{A}}_\pm.
\end{equation}
These superpositions for the fields $\bm{B}_h$ and $\bm{v}_h$ are found to be
\begin{align}
  &\bm{B}_h=\left( \frac{1}{f_1d_i}-d_i\lambda_+\right) \bm{\mathcal{A}}_+ +\left(\frac{1}{f_1d_i}-d_i\lambda_- \right) \bm{\mathcal{A}}_-\,, \label{eq:B_dB} \\
  &\bm{v}_h=\bm{\mathcal{A}}_+ + \bm{\mathcal{A}}_-\,, \label{eq:v_dB}
\end{align}
where the eigenvalues $\lambda_\pm$ are given by
\begin{equation}\label{eq:DB_beltrami_operator_eigenvalues}
    \lambda_\pm=\frac{-b_1\pm\sqrt{b_1^2-4b_2}}{2}\,.
\end{equation}
Therefore, the homogeneous solution to \eqref{eq:DB_matrix_nonhomogeneous} can be expressed in terms of the functions $\Psi_\pm$ satisfying
\begin{eqnarray}
    \Delta^* \Psi_\pm = - \lambda_{\pm}^2 \Psi_\pm\,, \label{GS_sigle_beltrami}
\end{eqnarray}
which stems from \eqref{eq:beltrami} with $\bm{\mathcal{A}}_\pm = R\mathcal{A}_{\phi}^{\pm}\bm{\nabla}\phi + \bm{\nabla} \Psi_\pm \times \bm{\nabla}\phi$. Equation \eqref{GS_sigle_beltrami} can be easily solved by separation of variables and one finds \cite{Cerfon_2014}:
\begin{equation}\label{eq:beltrami_flux_functions}
  \begin{aligned}
    &\Psi_\pm (R,Z)=R\sum\limits_k \biggr\{ {a_{k}}_\pm J_{1}\left(R\sqrt{\lambda_\pm^2-k^2}\right)\cos(kZ)+\\
    &+{b_{k}}_\pm Y_{1}\left(R\sqrt{\lambda_\pm^2-k^2}\right)\cos(kZ)\biggr\} + {c_1}_\pm R^2 \cos(\lambda_\pm Z) + \\
    &+{c_2}_\pm \cos(\lambda_\pm Z) + {c_3}_\pm \cos(\lambda_\pm\sqrt{R^2+Z^2}),
\end{aligned}
\end{equation}
with $J_1(x)$ and $Y_1(x)$ being the first order Bessel functions of the first and second kind respectively, and ${a_k}_\pm, \ {b_k}_\pm, \ {c_1}_\pm, \ {c_2}_\pm, \ {c_3}_\pm$ are unknown coefficients which will be determined in view of the boundary conditions.

Note that since the fields $(\bm{B}_h,\bm{v}_h)$ satisfy \eqref{eq:db}, which involves a double curl operator and are superpositions of two Beltrami fields, the corresponding states are called double Beltrami equilibrium states. Note additionally, that in the framework of MHD, the Beltrami states are homonymous to the well-known Taylor force-free states \cite{Cerfon_2014,marsh, wiegelmann_2012}, which are states with vanishing Lorentz force. 

To find the general solution to the nonhomogeneous system of elliptic partial differential equations \eqref{eq:DB_matrix_nonhomogeneous} we form a linear combination of the homogeneous solution expressed in terms of $\Psi_\pm$ and a particular solution of the inhomogeneous system. This general solution is
\begin{equation}\label{eq:psi_dB}
  \begin{aligned}
    \Psi(R,Z) = &\left( \frac{1}{f_1 d_i}-d_i \lambda_+ \right)\Psi_+ + \left( \frac{1}{f_1 d_i}-d_i \lambda_- \right)\Psi_- +\\ 
    &+\kappa_1 R^2 + \kappa_2,
  \end{aligned}
\end{equation}
\begin{equation}\label{eq:chi_dB}
  X(R,Z)= \Psi_+ +  \Psi_- +\kappa_3 R^2 + \kappa_4\,,
\end{equation}
Here $\kappa_j$, $j=\{1,...,4\}$ are the coefficients of the inhomogenous part of the solution and they are given by:
\begin{equation}\label{eq:DB_non_homogeneous_coefficients}
    \begin{aligned}
        & \kappa_1 = \frac{B_1 \mathcal{W}_4 -B_2 \mathcal{W}_2 }{\mathcal{W}_2 \mathcal{W}_3 - \mathcal{W}_1 \mathcal{W}_4}, \quad \kappa_2 =  \frac{A_1 \mathcal{W}_4 - A_2 \mathcal{W}_2}{\mathcal{W}_2 \mathcal{W}_3 - \mathcal{W}_1 \mathcal{W}_4}, \\
        & \kappa_3 = \frac{B_1 \mathcal{W}_3 - B_2 \mathcal{W}_1 }{\mathcal{W}_1 \mathcal{W}_4 - \mathcal{W}_2 \mathcal{W}_3}, \quad \kappa_4 = \frac{A_2 \mathcal{W}_1 - A_1 \mathcal{W}_3 }{\mathcal{W}_2 \mathcal{W}_3 - \mathcal{W}_1 \mathcal{W}_4},
    \end{aligned}
\end{equation}

\subsection{Equilibrium solutions in terms of Whittaker functions}

If one selects $g_1=-1/(d_i^2f_1)$ in the general ansatz \eqref{eq:ansatz1}--\eqref{eq:ansatz4}, then the system of GS equations \eqref{eq:first_gs}, \eqref{eq:second_gs} is decoupled and the two equations read as:
\begin{equation}\label{eq:Whit_non-homogeneous1}
  \begin{aligned}
    &\Delta^\star\Psi + (e_1 R^2 +e_2) \Psi + e_3 R^2 +e_4 =0 ,
  \end{aligned}
\end{equation}
\begin{equation}\label{eq:Whit_non-homogeneous2}
  \begin{aligned}
    &\Delta^\star\Phi + (\tilde{e}_1 R^2 + \tilde{e}_2) \Phi +\tilde{e}_3 R^2 +\tilde{e}_4 = 0.
  \end{aligned}
\end{equation}
 where the coefficients $e_j$, and $\tilde{e}_j$, ($j=1,...,4$) are given by
\begin{eqnarray}\label{eq:Whit_coefficients}
e_1 = n_2 \,, \quad e_2 = \left(\frac{1}{d_i^4f_1^2} - \frac{1}{d_i^2} \right)\,, \nonumber \\
e_3 = n_1, \quad e_4= -\frac{f_0+g_0}{d_i^2 f_1}\,, \nonumber \\
\tilde{e}_1 = -\frac{m_2}{d_i^2 f_1^2}\,, \quad \tilde{e}_2 = e_2 \,, \nonumber \\ 
\tilde{e}_3 = -\frac{m_1}{d_i^2 f_1^2}\,, \quad \tilde{e}_4 = e_4\,.
\end{eqnarray}
We can find analytic solutions to the homogeneous counterparts of the equations \eqref{eq:Whit_non-homogeneous1}--\eqref{eq:Whit_non-homogeneous2} employing the standard method of separation of variables. These homogeneous solutions depend on $R$ and on $Z$ through Whittaker functions \cite{whittaker_watson_1996} and trigonometric functions, respectively: 
\begin{equation}\label{eq:whit_homog_1}
  \begin{aligned}
    & \Psi_h(R,Z)=\sum\limits_k\biggr[a_k \mathscr{M}_{{\nu_k},\frac{1}{2}}(i\sqrt{e_1}R^2)\cos(kZ)+\\
    &+b_k \mathscr{W}_{{\nu_k},\frac{1}{2}}(i\sqrt{e_1}R^2)\cos(kZ) \biggr]\,,
  \end{aligned}
\end{equation}
\begin{equation}\label{eq:whit_homog_2}
  \begin{aligned}
    & \Phi_h(R,Z)=\sum\limits_k\biggr[\tilde{a}_k \mathscr{M}_{{\tilde{\nu}_k},\frac{1}{2}}(i\sqrt{\tilde{e}_1}R^2)\cos(kZ)+\\
    &+\tilde{b}_k \mathscr{W}_{{\tilde{\nu}_k},\frac{1}{2}}(i\sqrt{\tilde{e}_1}R^2)\cos(kZ) \biggr]\,.
  \end{aligned}
\end{equation}
Here $\mathscr{M}_{k,m}(z)$ and $\mathscr{W}_{k,m}(z)$ are the two independent Whittaker functions \cite{whittaker_watson_1996}, and%
\begin{equation}\label{eq:Whit_coefficients}
     \nu_k=i\frac{k^2-e_2}{4\sqrt{e_1}}, \quad \tilde{\nu}_k=i\frac{k^2-\tilde{e}_2}{4\sqrt{\tilde{e}_1}}
\end{equation}
with $i$ denoting the imaginary unit and $a_k, \ b_k, \ \tilde{a}_k, \ \tilde{b}_k$ are constant coefficients.

In the special case where the fractions $n_2/n_1$ and $m_2/m_1$ satisfy the following relation
\begin{equation}\label{eq:Whit_constraint}
   \frac{n_2}{n_1}=\frac{m_2}{m_1}=\frac{f_1-1/(d_i^2f_1)}{f_0+g_0},
\end{equation}
a set of two special solutions to the inhomogeneous equations \eqref{eq:Whit_non-homogeneous1}--\eqref{eq:Whit_non-homogeneous2} can be found by a direct similarity reduction method \cite{kaltsas_throum_2016} which is portrayed in the Appendix \ref{appendix}.  These solutions are
\begin{equation}\label{eq:whit_non_homog_1}
  \begin{aligned}
    &\Psi_p(R,Z)= \cos\left(\frac{\sqrt{e_1}}{2}R^2+\sqrt{e_2}Z\right)+\\
    &+\cos\left(\frac{\sqrt{e_1}}{2}R^2-\sqrt{e_2}Z\right)-\frac{e_3}{e_1}\,,
  \end{aligned}
\end{equation}
\begin{equation}\label{eq:whit_non_homog_2}
  \begin{aligned}
    &\Phi_p(R,Z)= \cos\left(\frac{\sqrt{\tilde{e}_1}}{2}R^2+\sqrt{\tilde{e}_2}Z\right)+\\
    &+\cos\left(\frac{\sqrt{\tilde{e}_1}}{2}R^2-\sqrt{\tilde{e}_2}Z\right)-\frac{\tilde{e}_3}{\tilde{e}_1}\,.
  \end{aligned}
\end{equation}
Since the partial differential equations \eqref{eq:Whit_non-homogeneous1}--\eqref{eq:Whit_non-homogeneous2} are linear, the general solutions will be constructed by superposing the homogeneous solutions \eqref{eq:whit_homog_1}--\eqref{eq:whit_homog_2} with the particular solutions \eqref{eq:whit_non_homog_1}--\eqref{eq:whit_non_homog_2}.

\section{Equilibria with shaped boundaries}\label{sec:4}

In this section we examine two applications of the equilibrium solutions of Section \ref{sec:3}, by Tokamak-relevant values for the various physical quantities and geometric parameters determining the plasma boundary. For simplicity, we have retained only cosine functions of $Z$, thus only up-down symmetric equilibria will be considered here. However, extension to up-down asymmetric equilibria is straightforward and will be pursued in future studies along with the construction of equilibria with negative triangularity and additional shape control features (e.g. \cite{Farengo_2020,Farengo_2020b}).

The solutions presented in the previous section contain a large number of unknown parameters that ought to be specified in view of appropriate boundary shaping conditions. For their determination, we will approximate a D-shaped boundary of interest with the following parametric equations
\begin{eqnarray}
  R_b(t) &=& 1 +\varepsilon\cos(\phi + \sin^{-1}\delta\sin t)\,, \nonumber \\
  Z_b(t) &=& \varepsilon\kappa \sin t\,, \label{eq:parametric_curve}
\end{eqnarray}
where $t\in[0,2\pi]$ is the poloidal angle, $\varepsilon$ is the inverse aspect ratio, $\kappa$ is the elongation, and $\delta=\sin\alpha$ is the triangularity \cite{cerfon_freidberg}. In terms of these geometrical parameters, we can select three characteristic points on the boundary with coordinates: $(1+\varepsilon,0)$ (outer equatorial point), $(1-\varepsilon,0)$ (inner equatorial point), and $(1+\delta\varepsilon,\kappa\varepsilon)$ (upper point). We may also define three curvatures \cite{cerfon_freidberg} in these points of interest
\begin{align}
  &N_1=\left[ \frac{d^2 R_b}{dZ_b^2}\right]_{\phi=0}=-\frac{(1+\alpha)^2}{\varepsilon\kappa^2}, \label{eq:curvature_1} \\
  &N_2=\left[ \frac{d^2 R_b}{dZ_b^2}\right]_{\phi=\pi}=\frac{(1-\alpha)^2}{\varepsilon\kappa^2}, \label{eq:curvature_2} \\
  &N_3=\left[ \frac{d^2 Z_b}{dR_b^2}\right]_{\phi=\pi/2}=-\frac{\kappa}{\varepsilon\cos^2(\alpha)}. \label{eq:curvature_3}
\end{align}
The boundary shaping conditions that we will employ for the determination of the unknown coefficients were first described in \cite{cerfon_freidberg} and subsequently utilized in a series of papers on Tokamak-pertinent analytic MHD equilibria \cite{Cerfon_2014,Throumoulopoulos_2012,Kaltsas_2014,  Evangelias_2016, Kaltsas_2019c}. These conditions are
\begin{align}
  &\Psi(1+\varepsilon,0)=0, \label{eq:boundary_1} \\
  &\Psi(1-\varepsilon,0)=0, \label{eq:boundary_2}\\
  &\Psi(1-\delta\varepsilon,\kappa\varepsilon)=0, \label{eq:boundary_3}\\
  &\Psi_R(1-\delta\varepsilon,\kappa\varepsilon)=0, \label{eq:boundary_4}\\
  &\Psi_{ZZ}(1+\varepsilon,0)=-N_1 \Psi_R(1+\varepsilon,0), \label{eq:boundary_5}\\
  &\Psi_{ZZ}(1-\varepsilon,0)=-N_2 \Psi_R(1-\varepsilon,0), \label{eq:boundary_6}\\
  &\Psi_{RR}(1-\delta\varepsilon,\kappa\varepsilon)=-N_3\Psi_Z (1-\delta\varepsilon,\kappa\varepsilon), \label{eq:boundary_7}
\end{align}
where the subscripts indicate partial derivatives with respect to the specified variable. Analogous shaping conditions can be considered also for $\Phi$. 

\subsection{Double Beltrami ITER-relevant equilibrium}

\begin{figure}[h]
	\centering
	\includegraphics[scale=0.35]{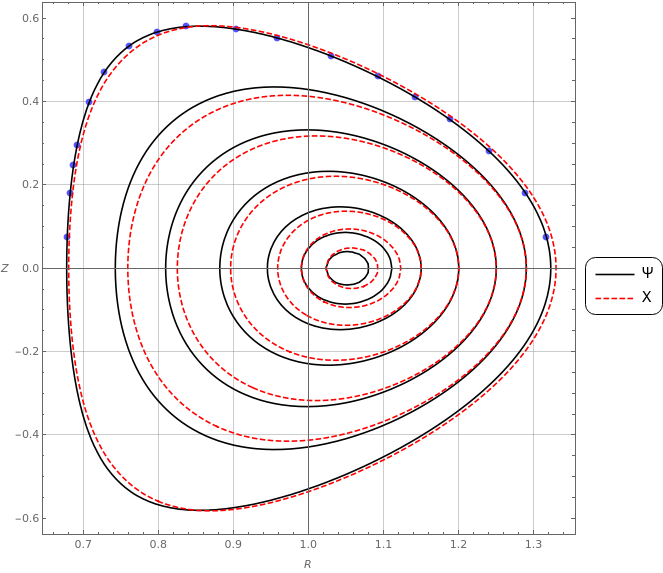}
	\caption{The contours of the magnetic and the ion velocity surfaces for the double Beltrami equilibrium. The positions of the extra points, selected for better shaping, are also marked.}
	\label{fig:DB_contours}
\end{figure}

On the basis of the solution \eqref{eq:beltrami_flux_functions}--\eqref{eq:chi_dB}, we constructed an up-down symmetric Tokamak equilibrium, with ITER-pertinent values for the geometrical parameters: $\varepsilon=0.32$, $\kappa=1.8$, $\delta=0.45$, $R_0=6.2\,m$,  and (normalized) ion skin depth $d_i=0.03$. The rest of the parameters were adjusted so that the equilibrium quantities attain large Tokamak values and profiles, although further optimization is possible. The plasma boundary was specified by Eqs.~\eqref{eq:boundary_1}--\eqref{eq:boundary_7}, along with some additional shaping conditions introduced to improve the boundary representation, where it was required that $\Psi$ vanishes at some additional boundary points. These conditions result in a linear system of algebraic equations which is solved numerically to determine the unknown coefficients ${a_k}_\pm, \ {b_k}_\pm$ ($k=1,...,5$) and  $\ {c_1}_\pm, \ {c_2}_\pm, \ {c_3}_\pm$ in \eqref{eq:beltrami_flux_functions} . Figure \ref{fig:DB_contours} depicts the nested magnetic and ion velocity surfaces in a poloidal cross-section, as well as the extra points selected on the boundary for improved shaping. This construction of double-Beltrami equilibrium with shaped, Tokamak-relevant boundary is an extension of a previous work \cite{Cerfon_2014}, concerned with exact Taylor states of toroidal, axisymmetric plasmas with Tokamak  geometric characteristics. The present equilibrium has some novel features due to its two-fluid nature, namely it accommodates strong plasma flows and pressure gradients with peaked pressure profile and closed isobaric surfaces and can describe high-beta equilibria. Furthermore, being a two-fluid equilibrium it contains additional information about the distinct electron and ion fluid behavior. We should note that the double-Beltrami states have been previously employed to study high-beta tokamak equilibria in \cite{Yoshida_2001}, however, the authors constructed those equilibria by means of numerical solutions on a domain with a simple circular boundary. 
It is remarkable that this simple analytic equilibrium model predicts a departure of the magnetic surfaces with the respect to the ion velocity ones, which is similar to the departure observed in previous works concerned with the calculation of numerical equilibria with compressible flows \cite{Kaltsas_2017, Guazzotto_2015}. The presented equilibrium configuration exhibits a positive Shafranov shift while it should be emphasised that due to the separation of the two fluids, there are two characteristic axes, a magnetic and a velocity one - though their positions differ slightly.

\begin{figure}[h!]
	\centering
	\includegraphics[scale=0.5]{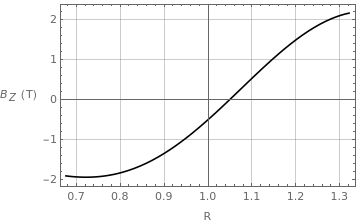}
     \includegraphics[scale=0.5]{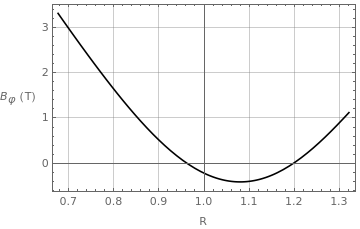}
	\caption{Magnetic field profiles for the double Beltrami equilibrium. Top: The $Z$-component of the magnetic field on the plane $Z=0$. Bottom: The toroidal component of the magnetic field on the plane $Z=0$.}
	\label{fig:DB_B_profiles}
\end{figure}

\begin{figure}[h!]
	\centering
	\includegraphics[scale=0.5]{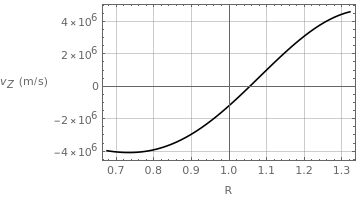}
    \includegraphics[scale=0.5]{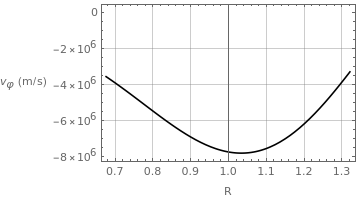}
	\caption{Ion velocity field profiles for the double Beltrami equilibrium. Top: The $Z$-component of the velocity field on the plane $Z=0$. Bottom: The toroidal component of the velocity field on the plane $Z=0$.}
	\label{fig:DB_v_profiles}
\end{figure}
\begin{figure}[h!]
	\centering
	\includegraphics[scale=0.55]{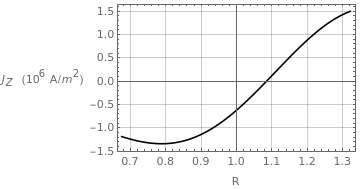}
    \includegraphics[scale=0.55]{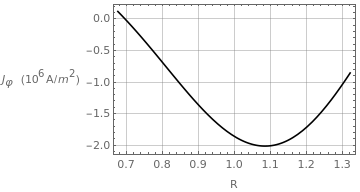}
	\caption{Current density profiles for the double Beltrami equilibrium. Top: The $Z$-component of the current density on the plane $Z=0$. Bottom: The toroidal component of the current density on the plane $Z=0$.}
	\label{fig:DB_J_profiles}
\end{figure}
\begin{figure}[h!]
    \centering
    \includegraphics[scale=0.5]{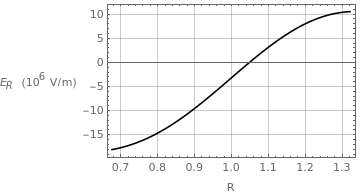} \includegraphics[scale=0.5]{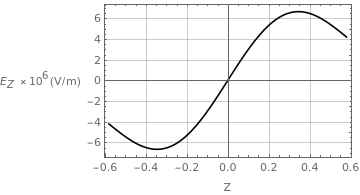}
    \caption{The $R$ and $Z$ components of the electric field on the $Z=0$ and $R=1$ planes respectively for the double Beltrami equilibrium.}
    \label{fig:DB_E_profiles}
\end{figure}
\begin{figure}[h!]
	\centering
	\includegraphics[scale=0.51]{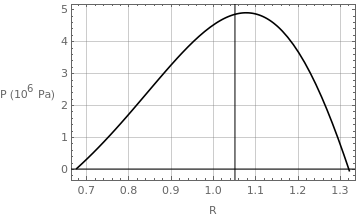}
	\caption{The plasma pressure profile on the $Z=0$ plane for the double Beltrami equilibrium.}
	\label{fig:DB_P_profile}
\end{figure}
\begin{figure}[h!]
	\centering
	\includegraphics[scale=0.35]{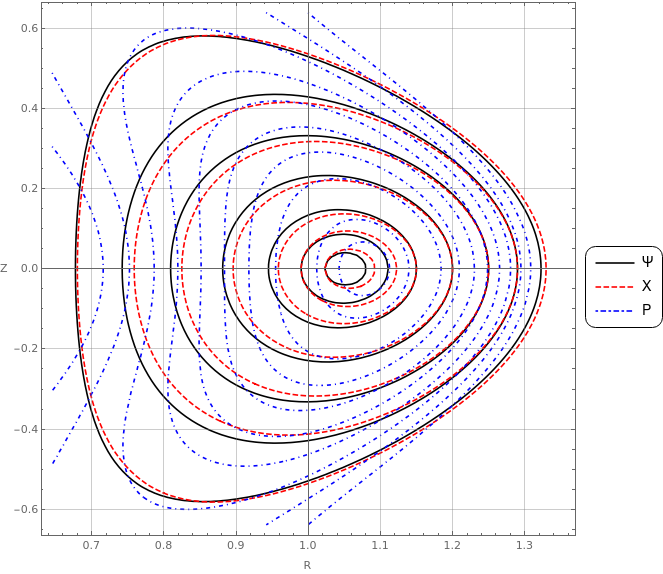}
	\caption{The pressure contours in a torus cross section, along with the magnetic and ion surfaces for the double Beltrami equilibrium.}
	\label{fig:DB_P_contours}
\end{figure}

In Figures \ref{fig:DB_B_profiles} and \ref{fig:DB_v_profiles} some profiles for the magnetic and ion velocity fields are illustrated. Both $\bm{B}$ and $\bm{v}$ have physically acceptable profiles, and their respective values are within acceptable ranges \cite{guazzotto, ernst}; the magnetic field values are $\mathcal{O}(1\,T)$, while the velocity values are $\mathcal{O}(10^6\, m/s)$. Moreover we observe that the toroidal magnetic field reverses in the core region, while the poloidal and toroidal velocity components have comparable magnitudes. Profiles of the current density components $J_Z$ and $J_\phi$ on the mid-plane $Z=0$ are displayed in Fig. \ref{fig:DB_J_profiles}. As we can see, the profiles are physically acceptable and the current density values are in the $MA/m^2$ range, which is typical for large Tokamaks, including ITER \cite{aymar_2002, sips_2005}. The $R$ and $Z$ components of the electric field, calculated by the Ohm's law \eqref{eq:ohm} are depicted in Fig.~\ref{fig:DB_E_profiles} presenting an expected behaviour and values on the order of $MV/m$.

Another equilibrium quantity that is of particular importance for the scope of magnetic confinement is the plasma pressure. Fig. \ref{fig:DB_P_profile} depicts a peaked-on-axis pressure profile for the double Beltrami equilibrium with typical values for large Tokamaks \cite{aymar_2002, sips_2014}. Owing to the flow the isobaric surfaces depart from the magnetic and the flow surfaces as it can be seen in Fig. \ref{fig:DB_P_contours}. As a result, although $P$ attains low values on the boundary, it is not exactly zero because the plasma flow does not vanish on $\partial \mathcal{S}$.

\subsection{ITER-relevant equilibrium in terms of the Whittaker functions}

\begin{figure}[h!]
	\centering
	\includegraphics[scale=0.27]{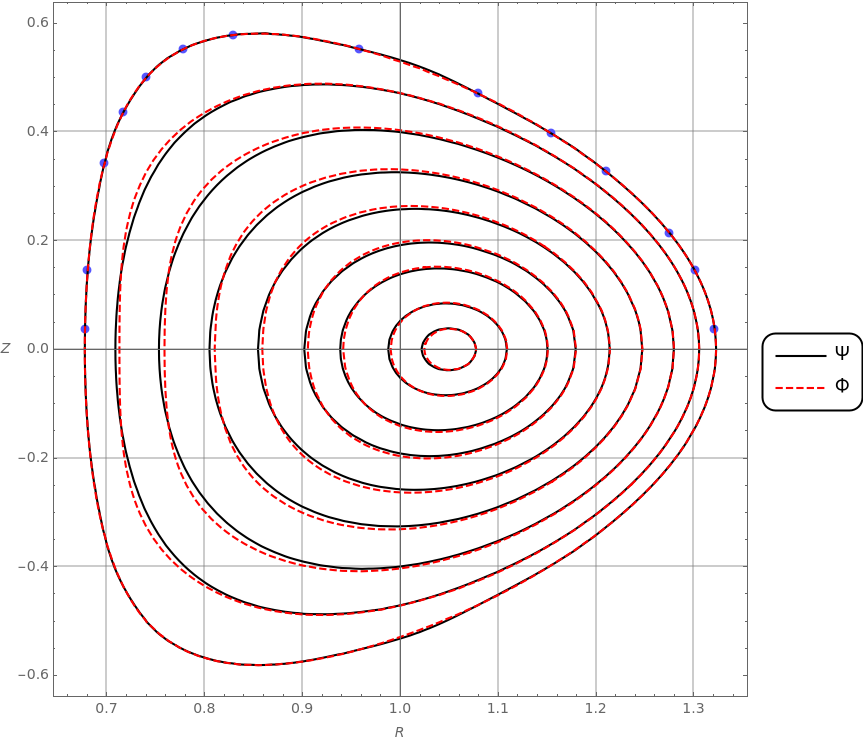}
	\caption{The contours of the magnetic and the ion velocity surfaces for the Whittaker equilibrium. The chosen extra points on the boundary are illustrated as well.}
	\label{fig:Whit_contours}
\end{figure}

This subsection is concerned with the construction of a second up-down symmetric Tokamak equilibrium, on the basis of the solutions \eqref{eq:whit_homog_1}--\eqref{eq:whit_non_homog_2}. From one point of view, this equilibrium is more general compared to the one constructed in the previous subsection since $m_2\neq 0$ and $n_2\neq 0$. The same ITER-relevant values for the geometric parameters were used as in the previous case, while for the ion skin depth we selected $d_i=0.02$. The rest of the parameters were again adjusted so that the equilibrium quantities approximate as much as possible Tokamak-relevant values and profiles. In this case the boundary shaping conditions were imposed simultaneously on both $\Psi$ and $\Phi$ since they are independent and as in the previous case some additional boundary points were introduced to improve the boundary representation. The coefficients $a_k, \ b_k, \ \tilde{a}_k, \ \tilde{b}_k$ where specified for $k=\{1,2,...,11\}$ this time. The nested contours of the two flux functions $\Psi$ and $\Phi$ are illustrated in Fig. \ref{fig:Whit_contours}.

As with the double Beltrami equilibrium, the separation of the two flux surfaces is evident. The velocity surfaces are organised around a distinct flow axis located very close to the magnetic one. A Shafranov shift of both surfaces can also be observed. A discernible difference with the previous equilibrium is that the two boundaries for $\Psi$ and $\Phi$ now coincide as we imposed the boundary on both $\Psi$ and $\Phi$.

\begin{figure}[h!]
	\centering
	\includegraphics[scale=0.38]{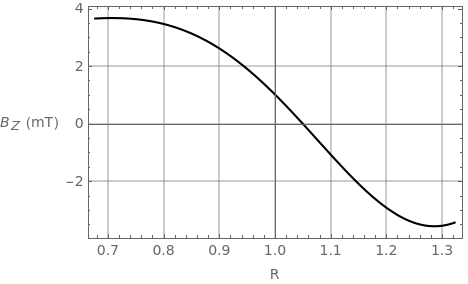}
 \includegraphics[scale=0.36]{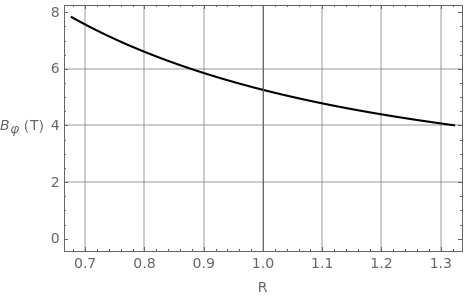}
	\caption{Magnetic field profiles for the Whittaker equilibrium. Top: The $Z$-component of the magnetic field on the plane $Z=0$. Bottom: The toroidal component of the magnetic field on the plane $Z=0$.}
	\label{fig:Whit_B_profiles}
\end{figure}
\begin{figure}[h!]
	\centering
	\includegraphics[scale=0.4]{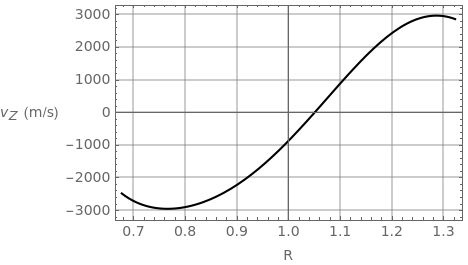}
 \includegraphics[scale=0.4]{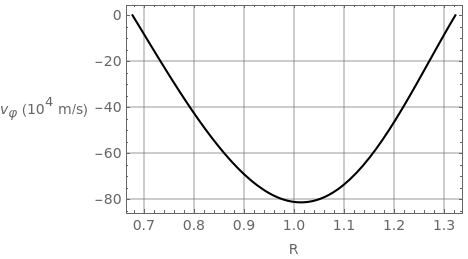}
	\caption{Ion velocity field profiles for the Whittaker equilibrium. Top: The $Z$-component of the velocity field on the plane $Z=0$. Right: The toroidal component of the velocity field on the plane $Z=0$.}
	\label{fig:Whit_v_profiles}
\end{figure}
\begin{figure}[h!]
	\centering
	\includegraphics[scale=0.39]{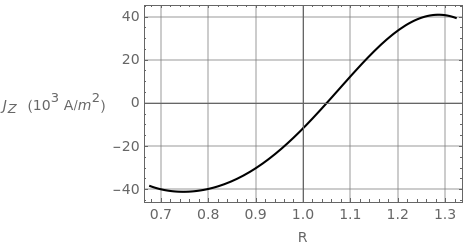}
 \includegraphics[scale=0.39]{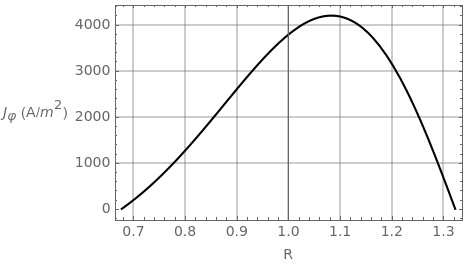}
	\caption{Current density profiles for the Whittaker equilibrium. Top: The $Z$-component of the current density on the plane $Z=0$. Bottom: The toroidal component of the current density on the plane $Z=0$.}
	\label{fig:Whit_J_profiles}
\end{figure}
\begin{figure}[h!]
	\centering
	\includegraphics[scale=0.36]{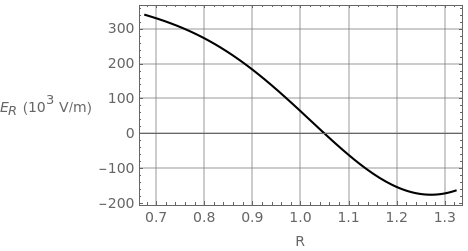}
    \includegraphics[scale=0.36]{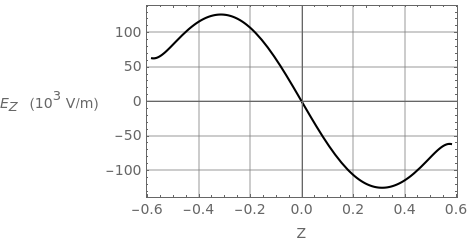}
	\caption{The $R$ and $Z$ components of the electric field on the planes $Z=0$ and $R=1$ respectively for the Whittaker equilibrium.}
	\label{fig:Whit_E_profiles}
\end{figure}
\begin{figure}[h!]
	\centering
	\includegraphics[scale=0.44]{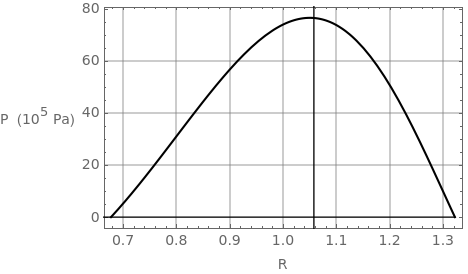}
	\caption{The plasma pressure profile on the $Z=0$ plane for the Whittaker equilibrium.}
	\label{fig:Whit_P_profile}
\end{figure}
\begin{figure}[h!]
	\centering
	\includegraphics[scale=0.27]{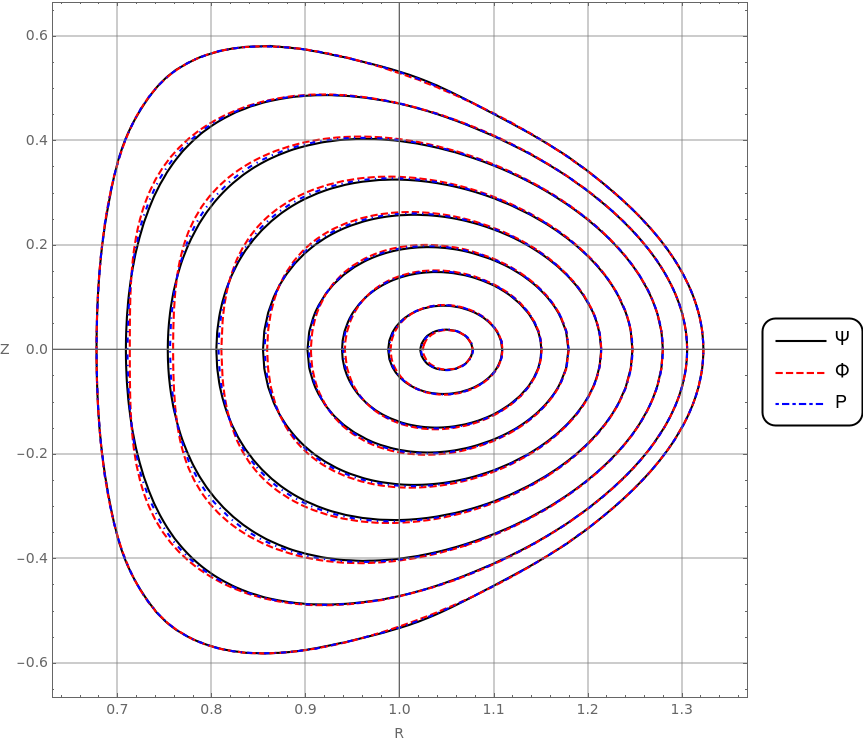}
	\caption{The pressure contours in a torus cross section, along with the magnetic and ion surfaces for the Whittaker equilibrium.}
	\label{fig:Whit_P_contours}
\end{figure}

In Fig. \ref{fig:Whit_B_profiles} we present the profiles of the $Z$- and $\phi$-components of the magnetic field on the mid-plane $Z=0$.  Although both profiles have a typical behavior and the toroidal magnetic field has desirable values, the poloidal magnetic field $\bm{B}_p$ attains particularly small values (of the order of $mT$). Regarding the velocity and the current density components, two velocity profiles are shown in Fig. \ref{fig:Whit_v_profiles}, exhibiting acceptable characteristics and two current density profiles are shown in Fig. \ref{fig:Whit_J_profiles}. Although both components exhibit  typical behavior, the toroidal current density, associated with the poloidal magnetic field, attains particularly small values. The small toroidal current of this particular equilibrium is insufficient to produce the rotational transform needed for effective confinement, so we tried to fix this inconsistency by  re-scaling some equilibrium parameters. However, this resulted in equilibria with unacceptably large values of other quantities (e.g. $P$), so we intend to employ a systematic procedure to further optimize the values of the free parameters for this particular class of equilibria in future work. 

Finally, we present the electric field components and the plasma pressure profile, calculated by means of Eq.~\eqref{eq:pressure} and \eqref{eq:ohm}, respectively. The results are shown, respectively in  Figs.~\ref{fig:Whit_E_profiles} and \ref{fig:Whit_P_profile}. The generated electric field ranges from $10^3$ to $10^4 \ V/m$, while the pressure has typical and desirable behavior as it peaks in the plasma core and almost vanishes on the boundary, with acceptable maximum values for large Tokamaks. As in the previous case, the isobaric surfaces form a distinct set of contours due to the flow contribution in the Bernoulli equation (Fig. \ref{fig:Whit_P_contours}) although the departure from the magnetic and flow surfaces is now less evident.

\section{Conclusions}\label{sec:5}

 The equilibrium GSB equations for axisymmetric, two-fluid plasmas with homogeneous ion density and anisotropic electron pressure  were derived using the energy-Casimir variational principle, stemming from the noncanonical Hamiltonian structure of Hall MHD. Subsequently, electron pressure anisotropy was omitted to enable the calculation of analytic equilibrium solutions. Two families of analytic solutions to the Hall MHD GS system were derived and employed to construct axisymmetric equilibrium states with Tokamak-pertinent characteristics.

The construction of the first type of equilibria (double Beltrami states), was motivated by previous work of the authors \cite{giannis} and was effected by superposing two Beltrami fields and a particular inhomogeneous solution. We demonstrated that this class of equilibria possesses Tokamak-relevant characteristics, namely nested magnetic and flow surfaces and acceptable values and profiles for the equilibrium quantities. Nevertheless, although the pressure in the double Beltrami equilibrium attains very low values on the boundary, it does not vanish thereon due to the nonvanishing flow.  In addition, there is considerable evidence \cite{gondal_2019} to postulate that double Beltrami states are essentially metastable equilibrium states, because, as Gondal et al. suggest, they tend to eventually relax to ordinary Beltrami states. This loss of equilibrium may take place under some circumstances, namely when certain scale parameters become degenerate or even when the product of the magnetic helicity with the ion helicity becomes positive \cite{gondal_2019, ohsaki_2002}. The abovementioned termination of the double-Beltrami equilibrium may also give rise to a conversion of magnetic energy to flow kinetic energy \cite{gondal_2019}. This metastability mechanism  suggests that double-Beltrami states can be candidates for the study of transient phenomena in laboratory plasmas but mainly in astrophysical environments \cite{ohsaki_2002, kagan_2010}.

Successive to the double Beltrami states, we studied a more general class of equilibria, with equilibrium solutions expressed in terms of Whittaker functions. The Whittaker equilibrium proved to be particularly interesting in some aspects, mainly because the pressure profile has a desirable behavior and optimal values. The rest of the quantities demonstrated typical profiles, with only exception the poloidal magnetic field and the toroidal current density values.  

For both classes of equilibria we employed a shaping method in terms of certain conditions that concern characteristic boundary points, allowing in this way the determination of the free parameters that appear in the analytic solutions. Both equilibrium classes can describe configurations with nested flux surfaces, peaked pressure profiles and isobaric surfaces which do not coincide with the other two sets of surfaces, i.e. the magnetic and the ion flow surfaces. Additionally, the non-parallel flow and the separation of the characteristic surfaces that characterizes both solutions, produces large-scale poloidal electric fields that may interfere in plasma and energy confinement and transport. It is intriguing that these analytic Hall MHD equilibria, recreate the general characteristics of the ion fluid and magnetic surface separation obtained in precedent numerical calculations \cite{Kaltsas_2017, Guazzotto_2015}. As a general remark we should note that this departure of the ion fluid from the magnetic surfaces, which is of the scale of the ion skin depth, is rather unimportant in terms of macroscopic equilibrium. However, such Hall corrections can be particularly important if these equilibria are used as reference states in the study of phenomena with scales relevant to the ion drift motions, like transport, microinstabilies, turbulence etc. 

In future work we intend to extend the present study by solving numerically the system \eqref{eq:first_gs}--\eqref{eq:second_gs} with electron pressure anisotropy ($\sigma \neq 0$) and further extend the model to incorporate ion pressure effects, anisotropy and finite Larmor radius corrections. Additionally, the analytic equilibria presented here could be extended to accommodate additional boundary shaping features, such as lower X-points and negative triangularity while a systematic determination of the free parameters in connection with the typical values of the physical quantities can also be pursued.

\section*{Acknowledgments}
This work has received funding from the National Fusion Program of the Hellenic Republic – General Secretariat for Research and Innovation.

\appendix

\section{Inhomogeneous solutions via direct similarity reduction}
\label{appendix}
The general solutions to the inhomogeneous equations \eqref{eq:Whit_non-homogeneous1}--\eqref{eq:Whit_non-homogeneous2} split into a homogeneous and an inhomogeneous part $u(R,Z) = u_h(R,Z) + u_p(R,Z)$, where $u_p=(\Phi_p,\Psi_p)$ is a particular, special solution of the corresponding inhomogeneous equation. The homogeneous part has been calculated in Section \ref{sec:3} in terms of Whittaker functions. On the other hand though, a special solution $u_p(R,Z)$ is hard to find. One can assume that $u_p(R,Z) = u(R)$ and thus the particular solution satisfies an ordinary differential equation of the form 
\begin{eqnarray}
    u_p''(R) - \frac{u_p'(R)}{R} + (e_1 R^2 + e_2)u_p + e_3 R^2 +e_4 =0\,,
\end{eqnarray}
where $e_j$, $j=\{1,...,4\}$ are constants. The general solution to this equation is again of the form $u_p(R)=w_h(R) + w_p(R)$ where $w_p(R)$ is a special inhomogeneous solution. We know \cite{Abell_2019} that in general the inhomogeneous part $w_p(R)$ can be calculated upon knowing $w_h(R)= c_1 h_1(R)+c_2 h_2(R)$  as follows:
\begin{eqnarray}
    w_p(R) = h_2(R) \int dR\, (e_3 R^2 + e_4)\frac{h_1(R)}{R} \nonumber\\
    - h_1(R) \int dR\, (e_3 R^2+e^4)\frac{h_2(R)}{R}\,,
\end{eqnarray}
where $h_1(R)$ and $h_2(R)$ are linearly independent solutions of the respective homogeneous equation.
Here, $h_1(R)$ and $h_2(R)$ are given in terms of Whittaker functions, and therefore one should resort in the use of numerical methods for the computation of the above integrals. For this reason, in this study, we employ a direct similarity reduction method \cite{kaltsas_throum_2016} which solves the system \eqref{eq:Whit_non-homogeneous1}--\eqref{eq:Whit_non-homogeneous2} in view of some constraints on the free parameters appearing in those equations. Both equations have the general form
\begin{eqnarray}
    \Delta^* u + (e_1 R^2 + e_2) u + e_3 R^2+ e_4 =0\,. \label{GS_general}
\end{eqnarray}
Let us consider particular solutions of the form
\begin{eqnarray}
u = u (\xi)\,, \quad \xi = c_R R^2 + c_Z Z\,, \label{part_sol_direct_red}
\end{eqnarray}
with $c_R$, $c_Z$ being constants. Substituting \eqref{part_sol_direct_red} into \eqref{GS_general} we can easily show that for $c_Z = \pm \sqrt{\frac{4 e_2}{e_1}} c_R$ and $ e_3 / e_4 = e_2/e_1$, the function $u(\xi)$ satisfies the following ordinary differential equation
\begin{eqnarray}
    4 c_R^2 u''(\xi) + e_1 u(\xi) + e_3 =0\,,
\end{eqnarray}
which is solved by
\begin{eqnarray}
    u(\xi) = c_1 \cos\left(\sqrt{\frac{e_1}{4 c_R^2}}\xi\right) \nonumber \\
+  c_2 \sin\left(\sqrt{\frac{e_1}{4c_R^2}}\xi\right)-\frac{e_3}{e_1}\,,
\end{eqnarray}
where $c_1$ and $c_2$ are arbitrary constants and $\xi= c_R\left( R^2 \pm \sqrt{\frac{4 e_2}{e_1}}\right) Z$. Employing this prescription for \eqref{eq:Whit_non-homogeneous1} and \eqref{eq:Whit_non-homogeneous2} and selecting $c_2=0$, we obtain the solutions \eqref{eq:whit_non_homog_1} and \eqref{eq:whit_non_homog_2} with the constraints \eqref{eq:Whit_constraint}.

\bibliographystyle{unsrtnat}
\bibliography{references} 

\end{document}